\documentstyle[preprint,prb,aps]{revtex}

\tighten
\begin{document}
\draft

\title{Hydrodynamic electron flow in high-mobility wires}

\author{M. J. M. de Jong\cite{byline} and L. W. Molenkamp\cite{byline2} }

\address{
Philips Research Laboratories,
5656 AA  Eindhoven,
The Netherlands
}

\date{Submitted October 24, 1994 --- {\tt cond-mat/9411067}}

\maketitle

\begin{abstract}
Hydrodynamic electron flow is experimentally observed
in the differential resistance
of electrostatically defined wires in the two-dimensional electron gas
in (Al,Ga)As heterostructures.
In these experiments
current heating is used to induce
a controlled increase in the number of electron-electron
collisions in the wire. The interplay between the
partly diffusive wire-boundary scattering and the electron-electron
scattering leads first to an increase and then to a decrease of
the resistance of the wire with increasing current.
These effects are the electronic analog of Knudsen and Poiseuille
flow in gas transport, respectively.

The electron flow is studied theoretically through a Boltzmann
transport equation, which includes impurity, electron-electron,
and boundary scattering.
A solution is obtained for arbitrary scattering parameters.
By calculation of flow profiles inside the wire it is demonstrated
how normal flow evolves into Poiseuille flow.
The boundary-scattering parameters for the
gate-defined wires can be deduced from the magnitude of the
Knudsen effect.
Good agreement between experiment and theory is obtained.
\end{abstract}


\narrowtext

\section{Introduction}
\label{s1}

In his 1909 paper on gas flow through a capillary, Knudsen demonstrated
that the ratio between the pressure drop over the capillary and the
gas-flow rate first increases and then decreases with increasing
density.\cite{knu09}
The mechanism is that with increasing density of gas particles, the
number of interparticle collisions increases. At low densities (what is now
known as the Knudsen transport regime) the gas particles
move almost independently, so that the flow is mainly
carried by particles with a large velocity parallel to the
wire axis. These particles travel long distances before colliding
with the wall. An occasional interparticle collision,
although not resistive by itself because of momentum conservation,
drives the parallel-moving particles towards the wall and shortens
their trajectories between subsequent collisions with the wall.
Therefore, in this regime, an enhancement of the
interparticle collision-rate leads to increasing dissipation of
forward particle momentum at the capillary walls.
At higher densities, however, many interparticle collisions
between subsequent particle-wall collisions occur,
resulting in a random-walk behavior.
As a consequence
a laminar (Poiseuille) flow evolves, in which the effective
particle-wall interaction is decreased.

Because of the analogy between classical diffusive transport of electrons
and gas particles, one anticipates that a similar transition between
Knudsen and Poiseuille flow may occur in electron transport.
In this case
electron-electron (e-e) scattering events are the analogue of collisions
between gas particles.\cite{Ziman}
Electron-electron scattering\cite{nees} has no influence on the
electrical resistivity of bulk materials,
because it conserves the total momentum of the electron distribution.
Effects of e-e scattering in the classical transport regime
can only be expected in the
resistivity of films and wires of high purity and small
dimensions,\cite{kav85} where conditions similar to those leading
to hydrodynamic gas flow can be realized.
Typically, the sample width $W$ should be smaller than or
comparable to the impurity mean free path $l_b$ of the bulk material.
These two lengths should be compared to $l_{ee}$, the
average length an electron covers between
two subsequent e-e scattering events.
When $l_{ee} > W$ one expects
an increase of the resistivity with increasing e-e scattering rate,
which is the electronic Knudsen effect.
In contrast,
when $l_{ee} < W$ the resistivity should decrease
with increasing e-e scattering rate, due to electronic Poiseuille flow.
The latter effect has been predicted by Gurzhi in 1963 \cite{gur63} and
is now known as the Gurzhi effect.
Experimentally, it proved difficult to obtain reliable data
on these effects,
because dissipation mechanisms not present in gas flow
usually prevent the occurrence of electronic Knudsen
and Gurzhi flow regimes:
First of all, electrons in a metal are scattered by impurities.
Moreover, since the e-e scattering rate
is usually varied by changing the lattice temperature of the sample,
the induced
effects are overwhelmed by electron-phonon interactions.
Furthermore, an increase in temperature also enhances the
{\em umklapp\/} electron-electron scattering rate, which adds
to the bulk resistivity.
Finally, deviations from an ideal spherical Fermi surface may
hinder interpretation of experimental data.

Due to these complications, only a few
experimental indications
of e-e scattering effects have been found.\cite{kav85}
Most experiments use potassium, as an exemplary simple
metal, which to a good approximation
has a spherical Fermi surface.\cite{bas90}
However, the observed changes in the
resistivity as a function of lattice temperature
are limited to about 0.01\% of the total resistivity,
because of the small $l_b$ and the onset of electron-phonon
scattering.
Yu {\em et al\@.} \cite{yu84} have reported a negative temperature
derivative of the resistivity ($ d \rho / d T$)
of potassium wires at temperatures around and below 1 K\@.
However, an interpretation in terms
of the Gurzhi effect was disputed,\cite{kav85}
since at these temperatures $l_{ee} > W$. In later publications of the
same group, it was shown that the negative $d \rho / d T$ can be
attributed to metallurgical imperfections,\cite{zha88}
and also Kondo-like effects in the resistivity were reported.\cite{yu89}
Observations of a
positive $d \rho / dT$ in wider wires \cite{zha88} were interpreted by
Movshovitz and Wiser \cite{mov90b} as a Knudsen-like behavior due to the
combination of e-e and
electron-phonon collisions. A similar mechanism was proposed
to explain an anomalously strong, positive $d \rho / d T$
in very thin potassium films.\cite{mov90a,qia91} However, until now there
has been no observation of electronic Poiseuille flow, nor has there
been an observation of a `Knudsen maximum' in the resistance\cite{Ziman} at
the crossover between Knudsen and Gurzhi flow regimes.

In this paper, we present an experimental and theoretical study
of Knudsen and Gurzhi transport phenomena in two-dimensional wires.
The wires used for the experiments are defined
electrostatically in the two-dimensional electron gas (2DEG) of
(Al,Ga)As heterostructures.\cite{C&H}
Using these devices to study hydrodynamic electron-flow
offers several advantages:
First, due to the high purity of the material and the resolution
of electron-beam lithography one can easily reach the condition
$l_b > W$.
Second, {\em umklapp\/} electron-electron scattering is absent,
because of the low electron density and the perfectly circular
Fermi surface.
Third, the electron-acoustic phonon
coupling is weak in the (Al,Ga)As-2DEG system.
This makes it possible to investigate the influence
of e-e scattering {\em not}
by changing the temperature $T$ of the full sample, but by selectively
changing the temperature $T_e$ of the electrons inside the
wire  by passing a dc current $I$ through the device.
Previously, this current-heating technique
has proven very useful for the study of
thermoelectric phenomena in nanostructures.\cite{gal90,mol90}
The wires studied here are equipped with
opposing pairs of quantum point-contacts in their boundaries. Since the
thermopower of the point contacts is quantized,\cite{mol90} we can
determine the electron temperature $T_e$ in the wire,
as a function of $I$, from a
thermovoltage measurement.\cite{mol92}
The ability to modify selectively
the e-e scattering rate allows a clear and unambiguous
demonstration of hydrodynamic effects on the resistance of the wire.

We measure in the experiments the differential resistance $dV/dI$ versus
$I$.\cite{diffR}
In the resistance curves we can
distinguish three regimes:
1) Starting from $I=0$ we observe an increase in $dV/dI$ with increasing
$I$. This is attributed to the Knudsen effect. We find resistance
changes as large as 10\% of the total resistance.
2) Then there is a range where $dV/dI$ decreases with increasing $I$,
which we identify as the Gurzhi effect.  In this range, we see relative
resistance changes up to 20\%.
3) Upon increasing $I$ we come into a regime where
$dV/dI$ increases again. Here, the heating due to the applied
current also affects the lattice temperature,
so that the resistance increase
can be attributed to enhanced electron-phonon scattering.
At the crossover between regime 1) and 2) the Knudsen maximum
is reached. The minimum in the resistance between regime 2) and 3)
was the actual subject of one of Gurzhi's first papers.\cite{gur63}

In order to understand our experimental results, we have developed
a theory based on the Boltzmann transport equation, which yields
quantitative agreement with the experiments.
In the first half of this century the Boltzmann approach has been
applied to study size effects on the resistance of small conductors.
The thin film case has been addressed by Fuchs \cite{fuc38} and
the case of a thin wire by Dingle.\cite{din50}
A particularly insightful method to solve the Boltzmann
equation is due to Chambers,\cite{cha50} who has expressed the
solution in terms of the effective mean free path the electron covers
between either bulk-impurity or boundary collisions.
These treatments consider partially diffusive boundary
scattering, in which part of the electrons colliding with the
boundaries is specularly reflected and the remainder is diffusely
scattered. The boundary scattering is modeled by a constant
specularity coefficient. In a more realistic treatment by
Soffer \cite{sof67} the wave nature of the electrons has been taken into
account and results in a specularity coefficient which depends on the
angle of incidence.
In Ref.\ \onlinecite{sam82} it is shown that inclusion
of the angle-dependent
specularity coefficient in a calculation of the resistivity of thin wires
gives a more satisfactory agreement
with experiments than Dingle's original theory.

The inclusion of e-e scattering in the Boltzmann
approach to the resistivity of wires
is not trivial and has been limited
to a certain parameter range in most treatments.
In the pioneering work by
Gurzhi,\cite{gur63} the situation $l_{ee} \ll l_b , W$ is considered.
It is shown that under these conditions the Boltzmann equation
can be mapped on a Navier-Stokes type of equation.
The opposite Knudsen regime $l_{ee} \gg l_b , W$ has been treated by
Movshovitz and Wiser,\cite{mov90b,mov90a} who use
the Chambers method to calculate
effective mean free paths with the
approximation that at most one e-e scattering event
in each electron trajectory is taken into account.
In Ref.\ \onlinecite{gur89} Gurzhi and coworkers provide an
alternative approach for this regime,
by solving the Boltzmann equation perturbatively.
This also allows including specific features of
the e-e scattering, such as the distinction between
isotropic and small-angle scattering.
We know of only two approaches that describe the resistivity of
wires from the Knudsen up to the Gurzhi regime. The first is due to
Black,\cite{bla80} who employs a Monte Carlo technique to calculate
effective mean free paths in a wire.
Although the numerical results are not so accurate because of
the limited computer power available at the time,
the Knudsen maximum in the resistivity is found.
The results show similar behavior for isotropic and small-angle
e-e scattering.
The second approach is due to De Gennaro and Rettori.\cite{gen84}
They start from the Boltzmann equation and include e-e scattering by
a scattering term due to Callaway \cite{cal59} in which the
electrons are relaxed towards a distribution with a net drift velocity.
As pointed out by Gurzhi {\it et al.},\cite{gur89}
the final results of Ref.\
\onlinecite{gen84} are incorrect,
because the spatial variation of the drift velocity is neglected.

Our theoretical description starts from a
kinetic equation similar to that of Ref.\ \onlinecite{gen84}.
We have obtained a self-consistent
solution of the relevant Boltzmann equation.
This is the first theory which provides an analytical expression
for the Boltzmann distribution function for {\it any} set of
$l_{ee}, l_b, W$.
It will prove insightful to express the Boltzmann distribution
function in terms of an effective mean free path.
For the regime $l_{ee} \gg l_b, W$ our solution
is equivalent to the results of Movshovitz and Wiser, so that we
have provided a formal basis for their method of including
e-e scattering events in the electron trajectories.
Our approach is indeed able to describe the transition from
Knudsen to Poiseuille flow. The transition can be illustrated by the
evolution of the electron-flow profiles along the wire.

In the three-dimensional case,
which has been addressed in most previous treatments,
the e-e scattering rate of electrons in a thermal slice around
the Fermi surface
is proportional to $T^2$, as follows
from the well-known phase space argument.\cite{Ziman}
For a 2DEG, instead, the e-e scattering rate
is proportional to $T^2 \ln T$.\cite{cha71,giu82}
In a study by Laikhtman \cite{lai92} of relaxation of injected electrons
into a zero-temperature 2DEG it is found that small-angle
scattering is important. Features of e-e
scattering in a 2DEG are also discussed by
Gurzhi and coworkers.\cite{gur94}
The e-e scattering term which we use is
first proposed by Callaway\cite{cal59} and
is not of a microscopic origin,
but takes the main feature of e-e scattering, conservation of momentum,
into account.
As we will show, an attractive feature of this simplified scattering term
is that it allows an exact (numerical) solution of the Boltzmann
equation.

We have compared experiment with theory
through a three step procedure:
First, using the results of the point-contact thermometry
we find $T_e$ versus $I$.
Then, using a formula due to Giuliani and Quinn\cite{giu82}
we calculate $l_{ee}$
as a function of $T_e$.
Finally, we determine the wire resistivity for the given $l_{ee}$
from our Boltzmann approach.
This has yielded quite a satisfactory agreement for both the
Knudsen and the Gurzhi regime. The regime 3) in which
phonon scattering due to the heating of the lattice
increases the resistivity is outside the range of our theory.
 From the magnitude of the Knudsen effect we obtain information on the
boundary-scattering parameters of the gate-defined wires.

A brief account of this work with an emphasis on the experiments
has already been published.\cite{mol94}
Here, we present a more extensive discussion.
Particular attention is paid to the
derivation of the theoretical model and how
its results can be compared with the experiments.
The outline of this paper is as follows:
In Sec.\ \ref{s2} the experiments are presented.
Sec.\ \ref{s3} describes the theoretical model
formulated in terms of a Boltzmann equation.
The method of solution and the
theoretical results including flow profiles are studied
in Sec.\ \ref{s4}.
Sec.\ \ref{s5} discusses the comparison between theory and experiment.
Finally, we conclude in Sec.\ \ref{s6}.
Appendix \ref{a1} and \ref{a2}
detail some technical parts of the calculation.

\section{Experimental observation of Knudsen and Gurzhi transport regimes}
\label{s2}

Our devices are fabricated from two different
(Al,Ga)As heterostructures containing a high-mobility 2DEG,
grown at Philips Research Laboratories, Redhill,
Surrey, UK. The wires used in the experiments are created by electrostatic
confinement of the 2DEG using a split gate technique. On top of the
heterostructures, which are mesa-etched in the shape of a Hall bar,
a pattern of TiAu gates is defined using electron-beam lithography.
The lay-out of the TiAu gates is given
schematically in the inset of Fig.\ \ref{f1}.
The wires have a lithographic width
$W_{\text{lith}} \simeq 4.0 \, \mu$m
(note that due to electrostatic depletion
the width $W$ of the wires in the 2DEG
is somewhat smaller),
and a length $L$ that varies between 20 and 120 $\mu$m.
A quantum point-contact\cite{C&H} is incorporated in each wire boundary.
We report here on three different types of samples,
whose particulars as to $L$, $W$, electron density $n$ and mean free path
$l_b$ are summarized in Table \ref{t1}.
For transport measurements, the samples are kept in a cryostat at
temperatures of 1.5 K and above, and at zero magnetic field. For reasons
of sensitivity, we measure the differential resistance of point
contacts and wires with standard low-frequency lock-in techniques,
using a 100 $\mu$V ac voltage.
All measurements are performed in a four-terminal geometry.

In order to adjust the electron temperature in the wires,
a dc heating current
$I \equiv I_{15} $
(typically an order of magnitude larger than the ac measuring current)
is passed through the wire using Ohmic contacts 1 and 5. Because of power
dissipation, the average kinetic energy of the electrons in the wire
increases. Due to frequent e-e scattering events, the
electron distribution-function
in the wire thermalizes rapidly to a heated
Fermi function at a temperature $T_e$, above the
lattice temperature $T$.
This increased electron temperature can be measured using the
quantum-point contacts in the wire boundaries:
since the electrons in the regions outside the wire remain at the
same temperature as the lattice,
a thermovoltage builds up across both
point contacts AB and CD, which can be measured as a transverse voltage
$V_{\rm trans} \equiv V_6-V_3$. Note that  $V_{\rm trans}$ does not
contain a contribution from the voltage drop along the wire, since point
contacts AB and CD face each other. We thus have
\begin{equation}
V_{\rm trans} \equiv V_6-V_3 = (S_{\rm AB}-S_{\rm CD}) (T_e-T)  \: ,
\label{e2.1}
\end{equation}
where $S_{\rm AB(CD)}$ denotes the thermopower of point contact AB(CD).

Like the electrical conductance,
the thermopower $S$ of a quantum point-contact exhibits a pronounced quantum
size-effect:\cite{str89,mol90} while the electrical conductance of the point
contact varies stepwise with the voltage on the split-gates, the thermopower
oscillates. The external gate voltage controls the number of one-dimensional
subbands present below the Fermi energy in the point contact. When the Fermi
energy inside the point contact falls in between two
subbands, the conductance is quantized, and the thermopower $S \simeq 0$.
However, when the Fermi energy inside the point contact exactly
coincides with
the bottom of the $N$-th subband, the conductance is in between
the $N$-th and the $(N-1)$-th plateau, and the thermopower attains a maximum
value, which for a step-function transmission probability of the point
contact,
is given by\cite{str89}
\begin{equation}
S=- \frac {k_B}{e}  \frac{\ln 2}{ N - \case{1}{2}} \: ,
\label{e2.2}
\end{equation}
if $N > 1$.
The quantum oscillations in the thermopower of a quantum point-contact
were predicted by Streda,\cite{str89} and an
experimental demonstration of the
effect has been reported elsewhere.\cite{mol90}
Here, we utilize the effect to
measure the electron temperature in the wire:
we adjust point contact CD on a conductance plateau,
thus setting $S_{\rm CD}
\simeq 0$, and adjust point contact AB for maximum thermopower
[$G_{\rm AB} = 1.5 \times (2e^2/h)$,
where $S_{\rm AB} \simeq - 40 \; \mu$V/K].
The result of such a measurement of $V_{\rm trans}$
as a function of dc heating
current $I$, obtained for a wire of type I, is shown in Fig.\ \ref{f1}.
For the longer wires a very similar behavior is found. In general,
we find that for $|I| \lesssim 20 \, \mu$A, and a lattice temperature
$T \lesssim 2$ K, the electron temperature $T_e$ in the wire
is approximately given by
\begin{equation}
T_e = T  + (I/W)^2 \sigma^{-1} \, C \: ,
\label{e2.3}
\end{equation}
where $\sigma$ is the conductivity of the wire.
The constant $C \simeq$ 0.05 m$^2$K/W\@.
Evidently, such a quadratic dependence of $T_e$ on $I$
is exactly what one expects to a first approximation for Joule dissipation.
For
$|I| \gtrsim 20 \, \mu$A, the situation is more complicated since at these
current levels also the lattice temperature starts to increase.

The hydrodynamic electron-flow effects that are the subject of this article
are observed in the differential resistance
$dV/dI \equiv dV_{\rm24}/dI_{\rm 15}$ of our
wires, as a function of dc heating current $I$.\cite{diffR}
Experimental results obtained for wires I, II, and III
for a series of lattice temperatures
are given in Figs.\ \ref{f2} and \ref{f3}.
Also shown are
theoretical results that will be
discussed in Sec.\ \ref{s5}. A strongly non-monotonic behavior of  $dV/dI$ is
evident for all traces.
This non-monotonic behavior in the differential resistance is the focus
of this paper and we will show that it results from electronic Knudsen and
Poiseuille flow.

A first remark we should make here is that in the
high-mobility 2DEG quantum
corrections to the resistance such as weak
localization are not measurable at the temperatures involved. This means
that the non-monotonic behavior must result from classical effects.
Note further that for the low lattice-temperature results of
Figs.\ \ref{f2} and \ref{f3}
all three resistance
regimes indicated in the Introduction can be observed:
1) Increasing $dV/dI$ due to Knudsen flow,
2) decreasing $dV/dI$ in the
Gurzhi regime,
and 3) a quasi-parabolically increasing $dV/dI$ due to lattice heating.
Only in the last regime, we find from a nearby thermometer that the lattice
temperature of the sample increases, implying that the
quasi-quadratic behavior 3)
is due to Joule heating of the lattice in combination
with the linear increase of electron-phonon scattering.\cite{kaw92}
Wire I (cf.\ Fig.\ \ref{f2})
exhibits a smaller Knudsen
effect (and only at the lowest lattice temperature studied) than
wires II and III\@.
As we will demonstrate
below, this results from the smaller ratio $l_b/W$ in wire I, compared to
wires II and III.
If the lattice temperature $T$ is increased we observe in Fig.\ \ref{f2}
two distinct effects. First, the $I=0$ resistance increases. This is
due to the decrease of $l_b$ by additional electron-phonon scattering.
Second, the hydrodynamic effects on the resistance disappear,
the Knudsen effect at lower $T$ than the Gurzhi effect.
This is caused by the decrease of $l_{ee}$ at $I=0$ (where $T_e=T$)
with increasing lattice temperature.
Another point to notice in Fig.\ \ref{f3} is that
the magnitude of the initial increase of
$dV/dI$ (the Knudsen effect) is twice as large for wire III
as for wire II. This shows that the effect scales
with the length of the wire and does not stem from e.g.\ the wire
entrances.

To see whether the hydrodynamic electron-flow phenomena mentioned in
Sec.\ \ref{s1} can indeed be responsible
for the anomalous behavior of $dV/dI$,
it is instructive to
estimate for wire I the e-e scattering mean free path  $l_{ee}$ for a
current $I=15  \, \mu$A, i.e.\ in the regime of decreasing $dV/dI$.
According to
Eq.\ (\ref{e2.3}), $I= 15 \, \mu$A corresponds to an electron temperature
$T_e \approx 16 \,$K (for a lattice temperature $T = 1.5 \,$K).
We have $l_{ee} = v_F \tau_{ee}$, where $v_F$ is the Fermi
velocity, and  $\tau_{ee}$  the e-e scattering time,
given by\cite{giu82,yac91,mol92b}
\begin{eqnarray}
   \frac{1}{\tau_{ee}} =\frac{E_F}{h}\left(
   \frac{k_B T_e}{E_F} \right)^2 \left[ \ln \left(
   \frac{E_F}{k_B T_e}\right) + \ln \left( \frac{2q}
   {k_F}\right) + 1 \right] \: .
\label{e2.4}
\end{eqnarray}
Here $q= m e^2/ 2\pi\varepsilon_{\rm r}\varepsilon_0\hbar^2$
is the Thomas-Fermi screening wavevector.
We find $l_{ee} \approx 0.8 \, \mu{\rm m}$,
which is much smaller than $W$. In this limit, the electrons undergo
a random-motion due to frequent e-e scattering events, and
we assign, at this stage tentatively,
the decrease in $dV/dI$ to the Gurzhi
effect. For currents below
$8 \, \mu$A, $dV/dI$ is positive. As
$l_{ee} \approx 5 \, \mu$m $\approx W$ for $I = 8 \, \mu$A and $T = 1.5\,$K,
the positive $dV/dI$ occurs in the right current range for the
electronic Knudsen effect.
In the following Sections we will formulate our calculations that
substantiate the assignment of the anomalous behavior
of $dV/dI$ to hydrodynamic electron flow.

\section{Boltzmann Equation}
\label{s3}

We study the electron flow inside a two-dimensional wire of width $W$
in response to a constant electric field $\bf E$, applied in the
$x$-direction, parallel to the wire.
The 2DEG has an ideal circular Fermi surface.
We look for a time-independent distribution function
$f({\bf r}, {\bf k})$ for electrons at position ${\bf r}= (x,y)$
and with wavevector ${\bf k}= k(\cos \varphi, \sin \varphi)$
(see inset of Fig.\ \ref{f1}),
which obeys the stationary Boltzmann transport equation
\begin{equation}
e {\bf E} \cdot
\frac{\partial f({\bf r}, {\bf k})}{\hbar \partial {\bf k}} +
{\bf v} \cdot \frac{\partial f({\bf r}, {\bf k})}{\partial {\bf r}} =
\left. \frac{\partial f({\bf r}, {\bf k})}{\partial t}
\right|_{\text{scatt}}
\: ,
\label{e3.1}
\end{equation}
where the r.\ h.\ s.\ is the scattering term, taking into account
both electron-impurity and e-e scattering.
Application of the electric field leads to a disturbance
of the distribution function from its equilibrium Fermi-Dirac
distribution $f_0(\varepsilon)=
1/\{ 1 + \exp[ ( \varepsilon - E_F )/k T_e] \}$
for energy $\varepsilon = \hbar^2 k^2 /2m=m v^2 /2$ and
with Fermi energy $E_F$.
At not too high fields, the non-equilibrium part of the
electron distribution function is only in a small shell
around the Fermi surface.
Therefore, and using the translational invariance along the
$x$-axis, we write the distribution function as
\begin{equation}
f({\bf r}, {\bf k}) = f_0(\varepsilon) +
\left( - \frac{\partial f_0}{\partial \varepsilon} \right)
\chi(y, \varphi) \: .
\label{e3.2}
\end{equation}
Substitution of Eq.\ (\ref{e3.2}) into Eq.\ (\ref{e3.1}) yields in linear
response
\begin{equation}
- e {\bf E} \cdot {\bf v} + {\bf v} \cdot {\bf \hat{y}}
\frac{\partial \chi(y,\varphi)}{\partial y} =
\left. \frac{\partial \chi(y,\varphi)}{\partial t}
\right|_{\text{scatt}}
\: ,
\label{e3.3}
\end{equation}
where $\bf \hat{y}$ is
the unit vector in the $y$-direction.
We neglect the energy dependence of the velocity in the thermal region
around the Fermi energy, so that
${\bf v}=v_F (\cos \varphi, \sin \varphi)$.

Once the distribution function has been evaluated, the current density
can be calculated according to
\begin{eqnarray}
{\bf j}(y) &=& 2 \sum_{\bf k} f({\bf r}, {\bf k}) e {\bf v} \: ,
\nonumber \\
&=& \int d \varepsilon {\cal D} (\varepsilon)
\left( - \frac{\partial f_0}{\partial \varepsilon} \right)
\int \limits_0^{2 \pi} \frac{d \varphi}{2 \pi}
\chi(y, \varphi) e {\bf v} \: , \nonumber \\
&=& e {\cal D} v_F \int \limits_0^{2 \pi} \frac{d \varphi}{2 \pi}
\chi(y, \varphi) {\bf \hat{v}}
\: ,
\label{e3.4}
\end{eqnarray}
with the two-dimensional density of states
${\cal D}(\varepsilon)={\cal D}=m/\pi \hbar^2$
(assuming a two-fold spin-degeneracy) and with unit vector
${\bf \hat{v}} = (\cos \varphi, \sin \varphi)$.

Let us now specify the scattering terms on the r.\ h.\ s.\ of Eq.\
(\ref{e3.3}).
The scattering by bulk impurities is
assumed to be elastic and isotropic. This implies for the
scattering term
\begin{equation}
\left. \frac{\partial \chi(y,\varphi)}{\partial t}
\right|_{b} = - \frac{\chi(y,\varphi)}{\tau_b} +
\frac{1}{\tau_b} \int \limits_0^{2 \pi} \frac{d \varphi'}{2 \pi}
\chi(y,\varphi') \: ,
\label{e3.5}
\end{equation}
where $\tau_b$ denotes the electron-impurity scattering time.
Note that the second term on the r.\ h.\ s.\ of Eq.\ (\ref{e3.5})
representing the electrons scattered {\em into} $(y,\varphi)$
is omitted in many treatments of the Boltzmann transport equation.
In these cases it is {\em a priori} assumed that the non-equilibrium
density is zero. For completeness, we maintain this term here and
show explicitly that it equals zero for our complete Boltzmann equation
in Appendix \ref{a1}.
For the e-e scattering term we
follow Refs.\ \onlinecite{gen84,cal59}
(the Callaway ansatz)
\begin{eqnarray}
\left. \frac{\partial \chi(y,\varphi)}{\partial t}
\right|_{ee} &=& - \frac{\chi(y,\varphi)}{\tau_{ee}} +
  \frac{1}{\tau_{ee}} \int \limits_0^{2 \pi} \frac{d \varphi'}{2 \pi}
\chi(y,\varphi')
\nonumber \\
&+& \frac{m {\bf v} \cdot {\bf v}_{\text{drift}}(y)}{\tau_{ee}}
\: ,
\label{e3.6}
\end{eqnarray}
with $\tau_{ee}$ the e-e scattering time and ${\bf v}_{\text{drift}}$
the net drift velocity. The e-e scattering term (\ref{e3.6})
implies that the electrons
are relaxed towards a
shifted distribution function
$f({\bf r}, {\bf k})=
f_0(\varepsilon - m {\bf v} \cdot {\bf v}_{\text{drift}})$.
The second term on the r.\ h.\ s.\ of Eq.\ (\ref{e3.6})
again ensures the conservation of particle density.
The drift velocity is related to the current density (\ref{e3.4})
 according to
${\bf j}(y) = n e {\bf v}_{\text{drift}}$ with the
electron density $n= {\cal D} E_F$, so that
Eq.\ (\ref{e3.6}) becomes
\begin{eqnarray}
\left. \frac{\partial \chi(y,\varphi)}{\partial t}
\right|_{ee} &=&
 - \frac{\chi(y,\varphi)}{\tau_{ee}}
\nonumber \\
&+& \frac{1}{\tau_{ee}}
\int \limits_0^{2 \pi} \frac{d \varphi'}{2 \pi}
\chi(y,\varphi') (1 + 2 {\bf \hat{v}'}\cdot {\bf \hat{v}})
\: .
\label{e3.7}
\end{eqnarray}
One readily verifies that this scattering term conserves the total
momentum
\begin{equation}
\int \limits_0^{2 \pi} d \varphi
\left. \frac{\partial \chi(y,\varphi)}{\partial t}
\right|_{ee} {\bf \hat{v}} = {\bf 0} \: .
\label{e3.7a}
\end{equation}
Actually, Eq.\ (\ref{e3.7}) is the simplest possible scattering term
with this property.
Since the scattering probability
from direction $\varphi$ to $\varphi'$ is proportional to
$1 + 2 {\bf \hat{v}'} \cdot {\bf \hat{v}} =
1 + 2 \cos(\varphi-\varphi')$,
small-angle forward scattering ($\varphi-\varphi' \approx 0$)
is most probable.
The negative values for $\varphi-\varphi' \approx \pi$
correspond to the scattering of
a non-equilibrium electron into
a non-equilibrium hole in the opposite direction.\cite{gur94}

For the scattering with the boundaries of the wire it is assumed
that a fraction $p$ of the incoming electrons is scattered specularly,
whereas the remainder is scattered diffusely.
In the original theories of size effects \cite{fuc38,din50,cha50}
the specularity coefficient $p$ is taken to be angle independent.
A microscopic model by Soffer \cite{sof67} for the scattering of the
incoming waves by the boundary roughness, finds that
$p$ depends on the angle of incidence
\begin{equation}
p(\varphi) = \exp[ - (  \alpha \sin \varphi )^2 ] \; .
\label{e3.8}
\end{equation}
This shows that electrons with grazing incidence
($\sin \varphi \rightarrow 0$) approach a unit
probability of specular reflection.
The parameter $\alpha = 4 \pi \delta / \lambda_{\rm F}$,
depends on the ratio between $\delta$, the root-mean-square
boundary-roughness, and the Fermi wavevector.

The boundary conditions for the solution of the Boltzmann
equation (\ref{e3.3}) are determined by
demanding particle conservation.
For the $y=0$ boundary we have
\begin{mathletters}
\label{e3.9}
\begin{eqnarray}
\chi(0,\varphi) &=& p(\varphi) \chi(0,2 \pi -\varphi)
\nonumber \\
&+&
\int \limits_{\pi}^{2 \pi} \frac{d \varphi'}{\pi}
[1 - p(\varphi')] \chi(0,\varphi') \: ,
\label{e3.9a}
\end{eqnarray}
if $\varphi \in [0,\pi]$, and for the $y=W$ boundary
\begin{eqnarray}
\chi(W,\varphi) &=& p(\varphi) \chi(W,2 \pi -\varphi)
\nonumber \\
&+& \int \limits_{0}^{\pi} \frac{d \varphi'}{\pi}
[1 - p(\varphi')] \chi(W,\varphi') \: ,
\label{e3.9b}
\end{eqnarray}
\end{mathletters}%
if $\varphi \in [\pi, 2 \pi]$.
The first term on the r.\ h.\ s.\ represents the specularly
reflected electrons, the second term the ones that are diffusely scattered.

To proceed, the non-equilibrium part of the distribution function is
written as \cite{cha50}
\begin{equation}
\chi(y,\varphi)=e E \cos \varphi \, l_{\text{eff}} (y,\varphi)
\: .
\label{e3.10}
\end{equation}
Here, the effective mean free path $l_{\text{eff}}(y,\varphi)$
can be interpreted
as the average length an electron at $y$ in the direction
$\varphi$ has covered
since the last boundary or impurity collision, as we show below.
It is clear that a replacement of
$l_{\text{eff}}(y,\varphi)$ in Eq.\ (\ref{e3.10})
by the bulk mean free path $l_b$ yields the well-known bulk solution
of the Boltzmann equation.
Let us now introduce mean free paths for bulk-impurity scattering
$l_b=v_F \tau_b$, for e-e
scattering $l_{ee}=v_F \tau_{ee}$, and for the combination of
those two $l^{-1}={l_b}^{-1} + {l_{ee}}^{-1}$.
As demonstrated explicitly in Appendix \ref{a1}, substitution
of Eq.\ (\ref{e3.10}) into the combined Eqs.\
(\ref{e3.3}), (\ref{e3.5}), and (\ref{e3.7}) gives
\begin{eqnarray}
&&\sin \varphi \frac{\partial l_{\text{eff}}(y,\varphi)}{\partial y}
+ \frac{l_{\text{eff}}(y,\varphi)}{l}
= 1 + \frac{\tilde{l}_{\text{eff}}(y)}{l_{\text{ee}}} \: ,
\label{e3.11} \\
&& \tilde{l}_{\text{eff}}(y) = \int \limits_{0}^{2 \pi}
\frac{d \varphi}{\pi} \cos^2 \! \varphi \: l_{\text{eff}}(y,\varphi) \: .
\label{e3.12}
\end{eqnarray}
The integro-differential equation (\ref{e3.11}) constitutes a major
simplification with respect to our starting point. This
result is the basis of our further analysis in the following Section.
The average effective mean free path $\tilde{l}_{\text{eff}}(y)$
is directly proportional to the drift velocity
\begin{equation}
{\bf v}_{\text{drift}}(y) = \frac{e {\bf E}}{m v_F}
\tilde{l}_{\text{eff}}(y)
\: ,
\label{e3.13}
\end{equation}
as follows from Eqs.\ (\ref{e3.4}), (\ref{e3.10}), and (\ref{e3.12}).
The conductivity of the wire, defined according to
${\bf j}= \sigma {\bf E}$, is given by
\begin{equation}
\sigma =  \frac{n e^2}{m v_F}
\int \limits_0^W \frac{dy}{W} \tilde{l}_{\text{eff}}(y)
=  \frac{n e^2}{m v_F}  L_{\text{eff}} \: .
\label{e3.14}
\end{equation}
The overall effective mean free path $L_{\text{eff}}$ is directly
proportional to the conductivity and will be used instead of $\sigma$
below.

\section{Theoretical results}
\label{s4}

As a preliminary application of Eq.\ (\ref{e3.11}) we briefly treat
the case of transport through a bulk conductor.
We thus seek a solution of the Boltzmann transport equation independent of
the spatial coordinates.
As a consequence of the disappearance of the $y$-dependence in
Eq.\ (\ref{e3.11})
it follows
that the effective mean free path $l_{\text{eff}}$ is independent of
$\varphi$ as well, so that [from Eq.\ (\ref{e3.12})]
$\tilde{l}_{\text{eff}}=l_{\text{eff}}$.
The solution of Eq.\ (\ref{e3.11}) is then easily found
\begin{equation}
l_{\text{eff}}=\frac{1}{l^{-1} - l_{ee}^{-1}} = l_b \: .
\label{e4.1}
\end{equation}
Note, that substitution into Eq.\ (\ref{e3.10}) produces the
ordinary bulk solution of the Boltzmann equation in the
absence of e-e scattering.
This solution is thus shown to be {\it independent} of the
e-e scattering rate.
It clearly demonstrates, that momentum-conserving e-e
scattering does not influence the bulk conductivity.

Let us now return to the wire, for which e-e scattering
can have a prominent influence on the conductivity.
As shown in Appendix \ref{a1}, it follows
from a symmetry argument that
$l_{\text{eff}}(y, \varphi) =l_{\text{eff}}(y, \pi - \varphi)$
for all $\varphi$.
It is then clear
from Eq.\ (\ref{e3.10}) that the
second term on the
r.\ h.\ s.\ of both Eqs.\ (\ref{e3.9a}) and (\ref{e3.9b})
vanishes.
The solution of Eq.\ (\ref{e3.11}) in combination with the
boundary conditions (\ref{e3.9})
can be written in the form of an
integral equation.
For clarity we first treat the case of completely diffusive boundary
scattering $p=0$. We then have for $\varphi \in [0, \pi]$
\begin{mathletters}
\label{e4.2}
\begin{eqnarray}
l_{\text{eff}}(y, \varphi) &=&
\int \limits_0^y \frac{dy'}{l \sin \varphi} \,
\frac{y - y'}{\sin \varphi} e^{-(y-y')/l \sin \varphi}
\nonumber \\
&+& \frac{y}{\sin \varphi} e^{-y/l \sin \varphi}
\nonumber \\
&+& \int \limits_0^y \frac{dy'}{l_{ee} \sin \varphi} \,
\tilde{l}_{\text{eff}}(y') e^{-(y-y')/l \sin \varphi}
\: ,
\label{e4.2a}
\end{eqnarray}
and for $\varphi \in [\pi, 2\pi]$
\begin{eqnarray}
l_{\text{eff}}(y, \varphi) &=&
\int \limits_y^W \frac{dy'}{l |\sin \varphi|} \,
\frac{y' - y}{|\sin \varphi|} e^{-(y'-y)/l |\sin \varphi|}
\nonumber \\
&+& \frac{W-y}{|\sin \varphi|} e^{-(W-y)/l |\sin \varphi|}
\nonumber \\
&+& \int \limits_y^W \frac{dy'}{l_{ee} |\sin \varphi|} \,
\tilde{l}_{\text{eff}}(y') e^{-(y'-y)/l |\sin \varphi|}
\: .
\label{e4.2b}
\end{eqnarray}
\end{mathletters}%
Eq.\ (\ref{e4.2}) elucidates the meaning of
the effective mean free path $l_{\text{eff}}(y, \varphi)$
as follows:
Each electron arriving at $y$
in the direction $\varphi$
has covered a certain path length
since the last diffusive scattering event.
The first term on the r.\ h.\ s.\ of Eq.\ (\ref{e4.2a})
takes into account the length covered from the last scattering event
at any $y'$ in between 0 and $y$. The exponential factor gives the
probability that the particle indeed reaches $y$ without any
additional scattering, whereas the distance covered is given by
$(y-y')/\sin\varphi$.
Note, that the scattering event at $y'$ might
have been either diffusive impurity scattering or e-e scattering.
In the latter case, also the path before the scattering event must
be accounted for, which is done by the last term. The second term denotes
the contribution of electrons after diffusive boundary scattering.
This interpretation of the solution of the Boltzmann equation
is originally due to Chambers.\cite{cha50}
The above derivation
demonstrates that this approach is still feasible when an
e-e scattering term is included in the Boltzmann
equation.
However, the solution itself is certainly more difficult to
obtain, since Eq.\ (\ref{e4.2}) must be solved self-consistently
with Eq.\ (\ref{e3.12}).

Previously, Movshovitz and Wiser have evaluated
the effect of e-e scattering on the resistivity of (three-dimensional)
films \cite{mov90a} and wires \cite{mov90b} by calculating
effective mean free paths with at most one e-e scattering event per
trajectory. This approach (most extensively described in
Ref.\ \onlinecite{mov90a})
yields valid results for the Knudsen regime
$l_{ee} \gg W, l_b$. We can treat this regime conveniently within our
formalism by
solving Eq.\ (\ref{e4.2}) perturbatively.
Only the result of the
first two terms of Eq.\ (\ref{e4.2}) is substituted into the third term.
One can prove that
this procedure is precisely
equivalent to that of Ref.\ \onlinecite{mov90a}.

In Appendix \ref{a2} we discuss a
perturbative analysis for the
two-dimensional wire with diffusive boundary scattering ($p=0$).
Here, we present the main results.
For the limit $l_b \gg W$ the conductivity [see Eq.\ (\ref{e3.14})]
in the absence of e-e scattering is given by
\begin{equation}
L_{\text{eff}} = \frac{2 W}{\pi}
\left[ \ln ( l_b / W ) + \ln 2 + \case{1}{2} - \gamma \right]
\: ,
\label{e4.3}
\end{equation}
where $\gamma$ is Euler's constant (see Appendix \ref{a2}).
In this limit the conductivity is directly proportional to
the width, whereas the dependence on the mean free path is
only present in the form of a logarithm.
The perturbative solution allows us to calculate the first
order correction to the conductivity due to e-e scattering.
For the situation $l_{ee} \gg l_b \gg W$ we find
\begin{equation}
\Delta L_{\text{eff}} =  - \frac{2 W l_b}{\pi l_{ee}}
\: .
\label{e4.4}
\end{equation}
We note that the conductivity {\em decreases} due to the
e-e scattering. This is the Knudsen effect. It is clear from
Eqs.\ (\ref{e4.3}) and (\ref{e4.4}) that the larger $l_b/W$ the
more prominent this effect becomes.
Previous calculations for this regime has yielded
$\Delta L_{\text{eff}}=-\case{3}{4} W l_b/l_{ee}$ for a three-dimensional
film of thickness $W$\cite{mov90a} and
$\Delta L_{\text{eff}} \sim - (W^2/l_{ee}) \ln(l_{ee}/W)$
for a three-dimensional
wire of diameter $W$.\cite{gur89}

For the opposite limiting regime $l_b \ll W$ the influence of the
boundary scattering on the conductivity becomes quite small.
 From the analysis in Appendix \ref{a2} we obtain
\begin{equation}
L_{\text{eff}} =  l_b -  \frac{4 l_b^2}{3 \pi W} \: .
\label{e4.5}
\end{equation}
The diffusive boundary scattering yields a small negative correction
to the bulk conductivity.
The first order influence of e-e scattering in the regime
$l_{ee} \gg W \gg l_b$ is
\begin{equation}
\Delta L_{\text{eff}} =  \frac{4 l_b^3}{15 \pi W l_{ee}}
\: .
\label{e4.6}
\end{equation}
Apparently, in this limit e-e scattering always
{\em increases} the conductivity, which
can be understood as follows: Since e-e scattering does not influence
the bulk conductivity, it can only change the small negative
correction due to the boundary scattering, represented by the second
term in Eq.\ (\ref{e4.5}).  Electron-electron
scattering decreases this correction,
which can be interpreted as the onset of the Gurzhi effect.
For comparison, we again mention results for three dimensions:
$\Delta L_{\text{eff}}= \case{6}{35}(\case{9}{8} - \ln 2) l_b^4/ W l_{ee}$
for a film (this can be calculated from the results given in Ref.\
\onlinecite{mov90a}) and
$\Delta L_{\text{eff}} \sim l_b^3/ W l_{ee}$
for a wire.\cite{gur89}

The calculation of the first order correction on the conductivity
due to e-e scattering thus displays an opposite behavior in the two limiting
regimes. This raises the question how $\Delta L_{\text{eff}}$ crosses over
from a positive value at small $l_b/W$ to a negative value at large
$l_b/W$.
One expects that the negative correction
to the conductivity appears when $l_b > W$. To substantiate
this expectation, we have calculated the correction for the
full regime of the ratio $l_b/W$. Details of this calculation
are given in Appendix \ref{a2}. The results are presented in
Fig.\ \ref{f4}, which depicts both the conductivity in the
absence of e-e scattering as well as the relative first order correction
due to e-e scattering as a function of $l_b / W$.
For the conductivity one observes a crossover from bulk-like
behavior [Eq.\ (\ref{e4.5})] to the logarithmic dependence
of Eq.\ (\ref{e4.3}). The first order correction in the conductivity
due to e-e scattering goes from a positive to a negative value.
We find that the Knudsen effect is only present for $l_b \gtrsim 1.3 W$.

The above results are valid for the regime of very low
e-e scattering rate. However, in order to compare with the experiments
we must also obtain solutions of Eq.\ (\ref{e3.11})
for the regime in which $l_{ee}$
becomes comparable with and smaller than $l_b, W$. In addition,
we need to incorporate the boundary condition (\ref{e3.9})
for arbitrary specularity coefficient $p(\varphi)$.
By transforming Eq.\ (\ref{e3.11}) into an integral equation
and integrating over $\varphi$ we find
\begin{equation}
\tilde{l}_{\text{eff}}(y) = \tilde{l}_{\text{eff}}^{(0)}(y)
+ \int \limits_0^W dy' G(y,y') \tilde{l}_{\text{eff}}(y') \: ,
\label{e4.7}
\end{equation}
\begin{eqnarray}
\tilde{l}_{\text{eff}}^{(0)}(y) &=& l - \frac{2 l}{\pi}
\int \limits_0^{\pi/2} d \varphi \, \cos^2 \! \varphi
\nonumber \\
&\times&
\frac{ [1-p(\varphi)]
\left[ e^{-y/l \sin \varphi} + e^{-(W-y)/l \sin \varphi} \right]}
{1 - p(\varphi) e^{-W/l \sin \varphi} }
\: ,
\label{e4.8}
\end{eqnarray}
\begin{eqnarray}
G(y,y')& =& \frac{2 }{\pi l_{ee}}
\int \limits_0^{\pi/2} d \varphi \,  \frac{\cos^2 \! \varphi}{\sin \varphi}
\Biggl\{ e^{-|y-y'|/l \sin \varphi} +
\nonumber \\
&&   \frac{ p(\varphi)
\left[ e^{-(y+y')/l \sin \varphi} + e^{-(2W-y-y')/l \sin \varphi} \right]}
{1 - p(\varphi) e^{-W/l \sin \varphi} }
\Biggr\}
\: .
\nonumber \\
&& \label{e4.9}
\end{eqnarray}
These are the key equations which allow the evaluation of the
conductivity for all values of $l_{ee}, l_b, W$, and $p$.
Essentially, the $\tilde{l}_{\text{eff}}^{(0)}$ term
is the two-dimensional equivalent of the Fuchs solution\cite{fuc38}
of the Boltzmann equation. The second term in Eq.\ (\ref{e4.7})
is a classical electron propagator-function which
takes the correction due to e-e scattering into account.
Note, that the perturbative approach as described in Appendix \ref{a2}
is equivalent to the approximation
$\tilde{l}_{\text{eff}}= (1 + G)\tilde{l}_{\text{eff}}^{(0)}$.
However, for larger values of $l_{ee}$ Eq.\ (\ref{e4.7}) must be
solved self-consistently according to
$(1-G)\tilde{l}_{\text{eff}} = \tilde{l}_{\text{eff}}^{(0)}$.
This can be achieved numerically
by discretizing the $y$-axis, so that Eq.\ (\ref{e4.7})
becomes a matrix equation.
This scheme allows the evaluation of
the solution $\tilde{l}_{\text{eff}}$ with a
precision which is only limited by the available
computer power.
We have used at least 400
gridpoints in our calculations to obtain sufficient precision.

In Fig.\ \ref{f5} the conductivity for a wire with diffusive
boundary scattering ($p=0$) is plotted against the
e-e scattering length for various values of the bulk-impurity mean free path.
For a wide wire ($l_b/W = 0.2$) the conductivity remains
approximately constant over the full range of $l_{ee}/W$.
The cases $l_b/W = 0.5, 1$ display a monotonous increase
of $L_{\text{eff}}$ with decreasing $l_{ee}$,
the Gurzhi effect.
Only for wires of width smaller than the mean free path ($l_b/W = 2, 5, 10$)
can both the Knudsen {\em and} the Gurzhi regimes be reached:
an initial decrease followed by an increase of
$L_{\text{eff}}$ with decreasing $l_{ee}$ is found from the calculation.
The Knudsen minimum in the conductivity is reached at
$l_{ee} \simeq W$.
It is clear that both the Knudsen effect
and the Gurzhi effect on the conductivity become
more prominent for larger ratios $l_b/W$.
We furthermore note that the conductivity saturates to its
bulk value ($L_{\text{eff}} \rightarrow l_b$) when the
e-e scattering rate becomes high ($l_{ee} \rightarrow 0$),
which reflects the vanishing influence of the boundaries
in this regime.

Let us now have a closer look at the effect of the
boundary scattering.
Fig.\ \ref{f6} displays the conductivity of a $l_b/W=5$
wire for various angle-independent specularity
coefficients $p$. The conductivity increases with
decreasing diffusive boundary scattering.
Besides this, we observe that for all $p<1$
both the Knudsen and the Gurzhi effect are found.
If the boundary scattering is fully specular ($p=1$),
$L_{\text{eff}} = l_b$ regardless of the amount of e-e scattering.
Essentially, the situation of specular boundary scattering
is equivalent to the bulk case, in
which the effects of e-e scattering are absent. It is easily checked that
$\tilde{l}_{\text{eff}}(y)=l_b$ solves Eq.\ (\ref{e4.7}) for $p=1$.
The relative conductivity change at the Knudsen maximum
$\Delta L_{\text{eff}} / L_{\text{eff}}$ (with respect to the
$l_{ee}=\infty$ value)
is depicted in the inset to Fig.\ \ref{f6}. It decreases
when the boundary scattering becomes less diffuse.

As we have remarked above, the modeling of the boundary
scattering by a constant specularity coefficient is only approximate.
Soffer\cite{sof67} has shown that a better
description is given by the angle-dependent specularity coefficient
of Eq.\ (\ref{e3.8}). Since the hydrodynamic effects in the
conductivity are caused by the interplay between the e-e scattering and the
boundary scattering, one may expect that the angle dependence
leads to differences in the magnitude of the Knudsen and Gurzhi effects.
Results comparing both models of boundary scattering
are shown in Fig.\ \ref{f7}.
The parameters in both models
are adjusted to yield equal
conductivity in the absence of e-e scattering.
It is clear from Fig.\ \ref{f7}
that the angle-dependent scattering leads to a much larger
Knudsen effect. The reason is as follows:
The conductivity is mainly determined
by electrons that move nearly parallel to the wire axis.
These electrons hit the boundaries at grazing incidence.
In the Soffer model electrons at grazing incidence experience a rather high
boundary specularity.
However, to have an equal
conductivity for both models in the absence of e-e scattering,
the boundary scattering of electrons with larger
incoming angles must be more diffusive in the Soffer model.
It is clear that this enhances the Knudsen effect.

So far, we have focused solely on
the conductivity. More insight in the microscopic
processes inside the wire can be
obtained from the solution $\tilde{l}_{\text{eff}}(y)$.
Since it is proportional to the drift velocity according to
Eq.\ (\ref{e3.13}), it represents the flow
profile across the wire.
Profiles for $l_b=5.5 W$ and $\alpha=0.7$ and various
amounts of e-e scattering
are shown in Fig.\ \ref{f8}.
In the absence of e-e scattering
 the drift velocity is almost constant as a function
of $y$.
On increasing the e-e scattering rate, the flow
profile over the full crosssection of the wire
shifts downwards due to the Knudsen effect:
Occasional e-e scattering events bend the electrons
moving parallel to the wire axis towards
the boundaries. This effectively decreases the drift velocity
and thus the conductivity.
However, for smaller $l_{ee}$ values
the flow profile develops a distinct curvature.
This indicates that electrons near the boundaries
experience more friction due to diffusive boundary scattering
than electrons in the middle of the wire.
The eventual result of this change in the flow profile
is that the conductivity increases with increasing e-e scattering rate,
the Gurzhi effect.
This behavior becomes more pronounced upon decreasing $l_{ee}$, and the
profile becomes similar to the classical, laminar Poiseuille flow.
Ultimately, however, the flow is limited by the bulk-impurity scattering,
as shown by the curve in Fig.\ \ref{f8}
for the smallest value of $l_{ee}$.
The electrons in the middle of the wire have a drift velocity
equal to the bulk value, whereas
close to the boundaries the drift velocity goes to zero.

In this Section
we have demonstrated which flow phenomena may occur in a
wire with both diffusive impurity scattering as well as
non-resistive e-e scattering.
In the next Section we present how the theory can be brought into agreement
with the experiments.

\section{Comparison between experiment and theory}
\label{s5}

Now that we have found that both the Knudsen
and the Gurzhi effect as observed in the experiments, cf.\ Sec.\ \ref{s2},
can at least qualitatively be understood by the theory
of the previous Sections,
we wish to make a more quantitative comparison.
Note, that the experimental
traces are $dV/dI$ versus $I$ curves, whereas the theoretical results
provide $L_{\text{eff}}$ as a function of $l_b, l_{ee}$, and $W$.

The resistance $R$ of the sample, as measured in the experiment,
is due to two contributions.
First, there is the resistance of the wire itself.
As shown in Ref.\ \onlinecite{jon94}, this is equal --- to a
good approximation ---  to the sum of the Drude resistance and
the Sharvin contact resistance.\cite{sha65}
The second contribution $R_0$ is due to
the unbounded regions in the 2DEG between the Ohmic contacts
and the entrance of the wire (see inset to Fig.\ \ref{f1}).
Note, that in an ideal four-probe measurement, the contacts should be
so close to the entrance of the wire, that this contribution
would be absent. In our samples, the typical distance
between the contacts and the wire is
on the order of 200 $\mu$m.
The actual value of $R_0$ may vary from wire to wire,
and with the lattice temperature.
 From previous experiments we estimate $R_0 \approx 60 - 90 \, \Omega$.
We thus have for the resistance\cite{jon94}
\begin{equation}
R= R_0 + \frac{h \pi}{2 e^2 k_F W} + \frac{L}{W \sigma} \: ,
\label{e5.1}
\end{equation}
in which the second term is the Sharvin resistance\cite{sha65}
and the third the Drude resistance.
The conductivity $\sigma$ is given by Eq.\ (\ref{e3.14}).
The values for $L, W, n$, and $l_b$ for each wire
are displayed in Table \ref{t1}.
Due to the electrostatic depletion, the
width $W$ of the wires is slightly smaller than the lithographic
width of the gate structure. For wire I we take $W=3.5 \, \mu$m and
for wires II and III $W=3.6 \, \mu$m.

The theoretical $L_{\text{eff}}$ versus $l_{ee}$ curve can now
be transformed into an $R$ versus $I$ curve in a three
step procedure.
First, we apply Eq.\ (\ref{e2.3}),
which gives the electron temperature $T_e$ against $I$.
Then, $l_{ee}$ is determined as a function of $T_e$
 through Eq.\ (\ref{e2.4}).
Finally, the Boltzmann theory provides
$L_{\text{eff}}$ (and thus $\sigma$) for the given $l_{ee}$,
so that the resistance is given by Eq.\ (\ref{e5.1}).
There is a little subtlety here, since
the resulting conductivity $\sigma$ is already used in Eq.\ (\ref{e2.3}).
One could adopt two approaches: The first would be to neglect the
dependence of $\sigma$ here
and simply use its $I=0$ value in Eq.\ (\ref{e2.3}).
The second approach, which we have applied, is to find a self-consistent
value of $\sigma$ and $l_{ee}$ in a numerical procedure.
Actually, this only slightly changes the $I$-axis.
 From the $R$ versus $I$ curve the differential
resistance $dV/dI$ versus $I$ is found.\cite{diffR}
It should be mentioned that
we do expect some deviations in the $I$-axis, because of
the approximate nature of Eq.\ (\ref{e2.3}).
Because of the limited validity of Eq.\ (\ref{e2.3}) we can only
treat the regime $|I| < 20 \mu$A\@.
This is sufficient since we only aim to model
the Knudsen and the Gurzhi regimes. The dissipative behavior
due to the heating of the lattice, which is
observed for higher currents in Figs.\ \ref{f2} and \ref{f3},
is not treated in the comparison.

In Fig.\ \ref{f9} we apply the above analysis
for the differential resistance of wire II at $T=1.8$ K\@.
The experimental curve is a blow-up of the
lowest temperature trace in Fig.\ \ref{f3}.
The theoretical curves are for various boundary-scattering parameters
and correspond to the plots in Fig.\ \ref{f7}
(since $l_b = 5.5 W$).
It should be stressed, that $R_0$ is not included in the
theoretical curves, since its precise value is not known.
This will be the case for all the comparisons.
Clearly the numerical results for a constant
specularity coefficient display a far too weak Knudsen and
Gurzhi behavior. Both effects can be increased by decreasing $p$,
but this also enhances the $I=0$ resistance to unreasonable values.
The plots in which the boundary scattering is taken to be
angle dependent --- using Eq.\ (\ref{e3.8}) --- display a much better
resemblance with the experiment.
Our experiments thus clearly indicate the validity of
Soffer's model\cite{sof67}
for boundary scattering in split-gate defined wires.
We find the best agreement with $\alpha=0.7$.
At $I=0$ the difference between the experimental and the theoretical
resistance
is 83 $\Omega$, which is within the right range of $R_0$.

We have applied the same analysis to the $T=1.5$ K
result of wire I\@.
As noted above, the magnitude of the Knudsen effect is much smaller
than in wires II and III
due to the lower ratio of $l_b/W = 3.5$.
This is indeed what is found in the theoretical calculation.
The comparison between theory and experiment is given
in the inset to Fig.\ \ref{f2}.
We have found that for wire I $\alpha=0.6$ yields the
best agreement.

The values of $\alpha$ that emerge from the comparisons
imply that the root-mean-square boundary
roughness of the gate-defined wires
$\delta \approx 2.5\,$nm  and that
approximately 80\% of the boundary scattering is specular.
This is consistent with earlier magneto-resistance
and electron-focusing experiments
in gate-defined 2DEG systems.\cite{C&H,tho89}
Note, that in the potassium-wires used for hydrodynamic
electron-flow experiments
the boundary scattering
is much more diffusive,
values of $\alpha \approx 25$ are used.\cite{mov90b}

Finally, we investigate the resistance behavior when
the lattice temperature is increased.
The experimental
curves for wire II and III for $T=$1.8, 3.5,
and 4.5 K are given in Fig.\ \ref{f3}.
The change in lattice temperature both influences
Eq.\ (\ref{e2.3}) as well as the bulk mean free path $l_b$,
which also
includes some electron-phonon scattering.
The difference in the $I=0$ resistance for the three temperatures
are thus caused by changes in $l_b$ and in
$l_{ee}$. Both increase the resistance with increasing
lattice temperature.
The decrease in $l_{ee}$ causes a part of the Knudsen correction
to be already incorporated in the $I=0$ value of $dV/dI$.
 From temperature-dependent mobility measurements we have
$l_b=18.5 \,\mu$m at $T=3.1$ K and $l_b=17.1 \,\mu$m at $T=4.5$ K\@.
Note, that for the theoretical analysis at $T=3.1, 4.5$ K
we push Eq.\ (\ref{e2.3}) slightly beyond its range
of validity.
A comparison with theory for $\alpha=0.7$
is presented in Fig.\ \ref{f3}.
For both wire II and wire III, the theoretical curves
are quite similar to
the experiments as to shape and amplitude.
The decrease in the Knudsen effect with increasing lattice temperature
is indeed found.
We do observe, however, a difference with the experiment for
the additional offset between the individual curves.
This is probably caused by a temperature dependence in $R_0$.

\section{Discussion and Conclusions}
\label{s6}

Our experiments have provided an unambiguous demonstration
of the occurrence of Knudsen and Gurzhi flow regimes in
electron transport.
The existence of these transport regimes has already been anticipated
in the 1960's.\cite{Ziman,gur63}
Although some aspects of hydrodynamic electron flow
have been observed
in potassium wires,\cite{yu84,zha88,yu89}
it is the high-mobility obtained in (Al,Ga)As heterostructures in
combination with nano-lithography techniques that has made the
observation of the complete transition from the Knudsen to the
Gurzhi flow regime accessible.
The current-heating technique appears to be an essential tool,
by which the e-e scattering rate can be varied, while keeping the
other types of scattering unaltered.
Due to the point-contact thermometry we
are able to determine the electron temperature inside the wire
as a function of the current.
Although hydrodynamic electron flow has been predicted many years ago,
its actual observation in our devices and the sheer size of the
effects is quite astonishing.

We have developed a theory based on the
Boltzmann transport equation.
The theory is more complex than that for
gas-flow because of the presence of bulk-impurity
scattering.
Most previous theoretical work\cite{gur63,mov90b,mov90a,gur89}
is only applicable to certain limiting flow regimes.
Our approach is more general, in the sense that it provides
the conductivity for the complete flow regime, i.e.\ for
any value of the wire width, the e-e scattering length,
and the bulk-impurity mean free path.
It should be mentioned that we have made two essential
simplifications in our Boltzmann approach.
First, we assume isotropic impurity scattering instead of the
small-angle scattering known to occur in a 2DEG.
Second, we apply a simple e-e scattering term due to Callaway,\cite{cal59}
which only takes into account
the conservation of the total momentum.
At this moment, we do not see a method of solution of
the Boltzmann equation
with on the one hand more realistic scattering terms,
and which is on the other hand applicable
to the complete transport regime.
However, our method already shows how complex the flow
behavior becomes due to the combination of resistive
impurity scattering
as well as partly diffusive boundary scattering and
non-resistive e-e scattering.

A quantitative comparison between experiment and
the Boltzmann theory
can be made, since the electron temperature
and thus the e-e scattering length inside
the wire can be inferred from experiment.
The obtained agreement is quite good.
This proves that in spite of its simplifications our
Boltzmann theory contains the essential physical ingredients to describe
the experiments.
Our results show that the Soffer model\cite{sof67} for
angle-dependent boundary scattering is more appropriate
to describe the scattering with the gate-defined wire boundaries
than a constant specularity coefficient.
Apart from the determination of the specularity
parameter, our comparison is only based on experimental data and
contains no fitting.

It would be of interest to perform further experiments
on hydrodynamic electron flow.
Promising areas of investigation are the influence of
more diffusive boundary
scattering, e.g.\ in wires defined by reactive ion etching
or ion exposure, and the application of a magnetic field.
The theoretical analysis given here can be adopted
in a straightforward manner to describe
the transition from Knudsen to Gurzhi flow in
three-dimensional systems.

\acknowledgements

We thank G. E. W. Bauer, C. W. J. Beenakker,
and H. van Houten for valuable discussions,
and E. E. Bende, O. J. A. Buyk, M. Kemerink, and
M. A. A. Mabesoone for technical assistance.
The heterostructures were grown by C. T. Foxon,
presently at the University of Nottingham, UK\@.
L. W. M. acknowledges
the kind hospitality of Y. Aoyagi, K. Ishibashi, J. Kusano, and S. Namba
during a visit to the Laboratory for Quantum Materials, RIKEN, Saitama,
Japan, where this research was initiated.
M. J. M. de J.
was supported by the ``Ne\-der\-land\-se
or\-ga\-ni\-sa\-tie voor We\-ten\-schap\-pe\-lijk On\-der\-zoek'' (NWO)
and by the ``Stich\-ting voor Fun\-da\-men\-teel On\-der\-zoek der
Ma\-te\-rie'' (FOM).

\appendix
\section{}
\label{a1}

We show how Eq.\ (\ref{e3.11}) can be derived.
The combination of the Boltzmann equation (\ref{e3.3}) with
the impurity (\ref{e3.5}) and the e-e (\ref{e3.7})
scattering terms yields
\begin{eqnarray}
&& - e E v_F \cos \varphi +
v_{F} \sin \varphi \,
\frac{\partial \chi(y,\varphi)}{\partial y} =
\nonumber \\
&& - \frac{\chi(y,\varphi)}{\tau} +
\frac{1}{\tau} \int \limits_0^{2 \pi}
\frac{d \varphi'}{2 \pi} \chi(y,\varphi')
\nonumber \\
&& + \frac{1}{\tau_{ee}}
\int \limits_0^{2 \pi} \frac{d \varphi'}{\pi}
\cos (\varphi-\varphi') \chi(y,\varphi')
\: ,
\label{eA1}
\end{eqnarray}
with $\tau^{-1}=\tau_b^{-1}+\tau_{ee}^{-1}$
For the time-independent case
the drift velocity has no component in the $y$-direction
\begin{equation}
\int \limits_0^{2 \pi} \frac{d \varphi}{\pi}
\sin \varphi \: \chi(y,\varphi) = 0
\: .
\label{eA2}
\end{equation}
As a result the $\cos (\varphi-\varphi')$ in the last term
in Eq.\ (\ref{eA1}) can be replaced by $\cos \varphi \cos \varphi'$.
Substitution of the parametrization (\ref{e3.10}) yields
\begin{eqnarray}
&& -\cos \varphi
+ \cos \varphi \, \sin \varphi \,
\frac{\partial l_{\text{eff}}(y,\varphi)}{\partial y} =
\nonumber \\
&& - \frac{\cos \varphi \, l_{\text{eff}}(y,\varphi)}{l}
+ \frac{1}{l} \int \limits_0^{2 \pi}
\frac{d \varphi'}{2 \pi} \cos \varphi' \, l_{\text{eff}}(y,\varphi')
\nonumber \\
&&+ \frac{\cos \varphi}{l_{ee}} \int \limits_0^{2 \pi}
\frac{d \varphi'}{\pi} \cos^2 \! \varphi' \, l_{\text{eff}}(y,\varphi')
\: .
\label{eA3}
\end{eqnarray}
Analysis of Eq.\ (\ref{eA3})  shows
that $l_{\text{eff}}(y,\varphi)$ and
$l_{\text{eff}}(y,\pi - \varphi)$ obey precisely the same
equation.
In addition the boundary conditions (\ref{e3.9}) are equal.
Due to this symmetry we have
\begin{equation}
l_{\text{eff}}(y,\varphi) = l_{\text{eff}}(y,\pi - \varphi)
\: .
\label{eA4}
\end{equation}
In combination with Eq.\ (\ref{e3.10}) it follows
that the non-equilibrium density is zero for all $y$
\begin{equation}
\int \limits_0^{2 \pi} d \varphi \, \chi(y,\varphi)
= e E \int \limits_0^{2 \pi} d \varphi \, \cos \varphi \,
l_{\text{eff}}(y,\varphi) = 0 \: .
\label{eA5}
\end{equation}
Thus, the second term on the r.\ h.\ s.\ of Eq.\ (\ref{eA3})
vanishes.
[This equally applies to the second terms
on the r.\ h.\ s.\ of Eqs.\ (\ref{e3.5}), (\ref{e3.7}), and (\ref{e3.9}).]
As a result Eq.\ (\ref{eA3}) leads to the
integro-differential equation (\ref{e3.11}) of the main text.

\section{}
\label{a2}

In this Appendix it is shown how some results presented in Sec.\ \ref{s4}
can be obtained.
We study the conductivity and its first order correction due
to e-e scattering
for a wire with diffusive boundary scattering ($p=0$).
By multiplication of Eq.\ (\ref{e4.2}) with $\cos^2 \! \varphi$
and integration over $\varphi$
one finds
\begin{eqnarray}
\tilde{l}_{\text{eff}}(y) &=&
l - \frac{2 l}{\pi} \int \limits_0^{\pi/2}
d \varphi \, \cos^2 \! \varphi \,
\left[ e^{-y/l \sin \varphi} + e^{-(W-y)/l \sin \varphi} \right]
\nonumber \\
&+& \frac{2}{\pi l_{ee}} \int \limits_0^W dy'
\int \limits_0^{\pi/2} d \varphi \,
\frac{\cos^2 \! \varphi}{\sin \varphi} \,
e^{-|y-y'|/l \sin \varphi} \tilde{l}_{\text{eff}}(y') \: .
\label{eB1}
\end{eqnarray}
In the limit of very small e-e scattering rate ($l_{ee} \gg l_b, W$) the
next step is to solve Eq.\ (\ref{eB1})
perturbatively. The first two terms
of Eq.\ (\ref{eB1}) are substituted into the third term.
An additional integration over $y$ then yields the conductivity
[cf.\ Eq.\ (\ref{e3.14})]
\begin{eqnarray}
L_{\text{eff}}&=& l - \frac{4 l^2}{\pi W} I(l/W)
+ \frac{l^2}{l_{ee}} - \frac{8 l^3}{\pi l_{ee} W} I(l/W)
\nonumber \\
&+& \frac{8 l^3}{\pi^2 l_{ee} W} K(l/W) \: ,
\label{eB2}
\end{eqnarray}
\begin{equation}
I(\lambda)= \int \limits_0^1 du \, u \sqrt{1-u^2}
\left( 1 - e^{-1/\lambda u} \right ) \: ,
\label{eB3}
\end{equation}
\begin{eqnarray}
K(\lambda) &=& \int \limits_0^1 du \, u \sqrt{1-u^2}
\int \limits_0^1 dv \, v \sqrt{1-v^2}
\nonumber \\
&\times&
\left[ \frac{1 - e^{-1/\lambda u-1/\lambda v}}{u+v} +
\frac{e^{-1/\lambda u}-e^{-1/\lambda v}}{u-v} \right] \: .
\label{eB4}
\end{eqnarray}
In the absence of e-e scattering ($l_{ee}=\infty$)
the conductivity is given by \cite{C&H}
\begin{equation}
L_{\text{eff}}= l_b - \frac{4 l_b^2}{\pi W} I(l_b/W) \: .
\label{eB5}
\end{equation}

The first order correction due to e-e scattering
can be found by substracting
Eq.\ (\ref{eB5}) from (\ref{eB2}) and expanding
$l=l_b - l_b^2/l_{ee}$.
The result can be evaluated analytically in two limits.
For a very wide wire $l_b \ll W$
we use the results
\begin{mathletters}
\begin{eqnarray}
\lim \limits_{\lambda \rightarrow 0} I(\lambda) &=& \frac{1}{3} \: ,
\label{eB6a} \\
\lim \limits_{\lambda \rightarrow 0} K(\lambda) &=& \frac{\pi}{30} \: ,
\label{eB6b}
\end{eqnarray}
\end{mathletters}%
which provide Eqs.\ (\ref{e4.5}) and (\ref{e4.6}).
In the opposite limiting regime of a very narrow wire ($l_b \gg W$)
the integrals (\ref{eB3}) and (\ref{eB4}) are more complex.
We have obtained the following series expansions
\begin{mathletters}
\begin{eqnarray}
\lim \limits_{\lambda \rightarrow \infty} I(\lambda) &=&
\frac{\pi}{4 \lambda} - \frac{1}{2 \lambda^2} ( \ln 2\lambda
+ \case{1}{2} - \gamma) + {\cal O}
(\lambda^{-3})
\: ,
\label{eB7a} \\
\lim \limits_{\lambda \rightarrow \infty} K(\lambda) &=&
\frac{\pi^2}{8 \lambda} - \frac{\pi}{2 \lambda^2} ( \ln 2\lambda
+ \case{1}{2} - \gamma) + {\cal O}
(\lambda^{-3})
\: ,
\label{eB7b}
\end{eqnarray}
\end{mathletters}%
where $\gamma \simeq 0.577$ is Euler's constant.
These results yield Eqs.\ (\ref{e4.3}) and (\ref{e4.4}).

The first order correction due to e-e scattering
in between these two regimes
can be evaluated by substracting Eq.\ (\ref{eB5}) from (\ref{eB2}).
We then have
\begin{equation}
\Delta L_{\text{eff}} = \frac{l_b}{l_{ee}} \left[
\frac{8 l_b^2}{\pi^2 W} K(l_b/W) - \frac{4 \l_b}{\pi} J(l_b/W) \right]
\: ,
\label{eB8}
\end{equation}
\begin{equation}
J(\lambda)= \int \limits_0^1 du \sqrt{1-u^2} \,
e^{-1/\lambda u}  \: .
\label{eB9}
\end{equation}
By numerical integration of $I$, $J$, and $K$ the plots in Fig.\ \ref{f4}
are obtained.

\begin{figure}
\caption{
Dependence of the thermovoltage
$V_{\rm trans}\equiv V_6-V_3$ and of the difference between the electron
and the lattice temperature $T_e-T$
on the heating current $I$ measured for wire
I at $T=1.5$ K\@.
Point contact AB is adjusted for maximum, CD for zero thermopower.
Inset:
Schematical layout of the gates (hatched areas) used to define a wire
with point-contact
voltage probes. The wire width $W$ is typically 4 $\mu$m,
the length $L$ varies between 20 and 120 $\mu$m. The crossed boxes denote
ohmic contacts. The coordinates used for the theory are indicated.
}
\label{f1}
\end{figure}

\begin{figure}
\vspace{0.5cm}
\caption{ Differential resistance $dV/dI$ of wire I
as a function of current $I$ for lattice
temperatures $T=$
24.7, 20.4, 17.3, 13.6, 10.4, 8.7, 4.4, and 1.5 K (from top to bottom).
The upper panel (a) of the inset is a magnification of the $T=$1.5 K result.
The lower panel (b) displays the
result of the theory described in
Sec.\ \protect\ref{s5}.
}
\label{f2}
\end{figure}

\begin{figure}
\vspace{0.5cm}
\caption{ Differential resistance $dV/dI$ vs.\ current $I$ for wire II and
III for lattice temperatures of (from top to bottom)
$T=$4.5, 3.1, and 1.8 K\@. At higher current levels, $dV/dI$ exhibits a
quasi-quadratic increase with current, similar to that in
Fig.\ \protect\ref{f2}.  Left panel (IIa) and (IIIa): experimental
traces; right panel (IIb) and (IIIb):  results of calculations,
see Sec.\ \protect \ref{s5}.}
\label{f3}
\end{figure}

\begin{figure}
\vspace{0.5cm}
\caption{
The conductivity $L_{\text{eff}}$ in the absence of e-e scattering
and the first order correction $\Delta L_{\text{eff}}$ due
to e-e scattering against the bulk-impurity mean free path $l_b$.
Results are for a two-dimensional wire
with diffusive boundary
scattering ($p=0$) according to
Eqs.\ (\protect\ref{eB5}) and (\protect\ref{eB8}), respectively.
}
\label{f4}
\end{figure}

\begin{figure}
\vspace{0.5cm}
\caption{
The conductivity $L_{\text{eff}}$ of a
wire with diffusive boundary scattering ($p=0$)
against the e-e scattering mean free path $l_{ee}$ for various
bulk-impurity mean free paths $l_b$.
}
\label{f5}
\end{figure}

\begin{figure}
\vspace{0.5cm}
\caption{
The conductivity $L_{\text{eff}}$ of a
wire with a mean free path $l_b=5 W$
against the e-e scattering mean free path $l_{ee}$ for various
specularity coefficients $p$.
The inset shows the relative change in the conductivity
at the Knudsen maximum
(which corresponds to the {\em minimum} in the conductivity).
}
\label{f6}
\end{figure}

\begin{figure}
\vspace{0.5cm}
\caption{
Comparison of the conductivity $L_{\text{eff}}$
as a function of $l_{ee}$
for constant boundary-scattering
coefficients (dotted curves) and for
angle-dependent coefficients (solid lines)
according to Eq.\ (\protect\ref{e3.8}).
To have approximately equal conductivity in the absence of e-e scattering
the comparison is between (top to bottom)
$p=0.895, 0.87, 0.845$ and
$\alpha=0.6, 0.7, 0.8$, respectively.
The bulk-impurity mean free path $l_b=5.5 W$.
}
\label{f7}
\end{figure}

\begin{figure}
\vspace{0.5cm}
\caption{
Velocity profiles inside the wire show how the flow
changes from the Knudsen upto the Gurzhi regime.
Depicted are the (normalized)
drift velocity $\tilde{l}_{\text{eff}}(y)$ as a function of
the transverse coordinate $y$ for $l_{ee}/W=$  100 ($\times$),
1 ($\triangle$),
0.1 ($+$),  0.01 ($\Box$), and 0.001 ($\Diamond$).
The inset shows the conductivity $L_{\text{eff}}$
as a function of the e-e scattering length
$l_{ee}$ and the symbols that indicate to which value
each flow profile corresponds.
Results are for the bulk mean free path $l_b=5.5 W$ and
for angle-dependent boundary scattering with $\alpha=0.7$.}
\label{f8}
\end{figure}

\begin{figure}
\vspace{0.5cm}
\caption{
Differential resistance $dV/dI$ versus current $I$ for wire II\@.
The top curve is the experimental result at $T=1.8$ K, as
shown for a larger current range in Fig.\ \protect\ref{f3}.
The other curves are theoretical results for various
boundary-scattering parameters. The dotted lines are
calculated with a constant specularity coefficient
$p=0.845, 0.87, 0.895$ (top to bottom).
The solid lines are calculated
for angle-dependent boundary scattering,
with $\alpha=0.8, 0.7, 0.6$ (top to bottom).
Best agreement with experiment is found for $\alpha=0.7$
(thick curve).}
\label{f9}
\end{figure}

\begin{table}

\caption{ Length $L$, lithographic width $W_{\text{lith}}$,
electrical width $W$,
electron density $n$, mean free path
$l_{\rm b}$ [at 1.5 K (sample I) and 1.8 K (sample II \& III)],
and specularity parameter $\alpha$ of the samples discussed in this paper.}
\begin{tabular}{cdddddd}
\multicolumn{1}{c}{Sample} &
\multicolumn{1}{c}{$L$} &
\multicolumn{1}{c}{$W_{\text{lith}}$} &
\multicolumn{1}{c}{$W$} &
\multicolumn{1}{c}{$n$ } &
\multicolumn{1}{c}{$l_b$} &
\multicolumn{1}{c}{$\alpha$} \\
\multicolumn{1}{c}{} &
\multicolumn{1}{c}{($\mu$m)} &
\multicolumn{1}{c}{($\mu$m)} &
\multicolumn{1}{c}{($\mu$m)} &
\multicolumn{1}{c}{($10^{11}$cm$^{-2}$)} &
\multicolumn{1}{c}{($\mu$m)} &
\multicolumn{1}{c}{} \\
\hline
  I     &  20.2  & 3.9 & 3.5 & 2.2 & 12.4  & 0.6 \\
  II    &  63.7  & 4.0 & 3.6 & 2.7 & 19.7  & 0.7 \\
  III   &  127.3 & 4.0 & 3.6 & 2.7 & 19.7  & 0.7 \\
\end{tabular}
\label{t1}

\end{table}


\begin{references}
\bibitem{diffR}
The relation between the resistance and the differential resistance
at a current $I$ is: $dV/dI  = R + I dR/dI$.
\end{references}
\end{document}